\documentclass[preprint,showpacs,preprintnumbers,amsmath,amssymb,onecolumn,superscriptaddress]{revtex4}

	\usepackage{graphicx,dcolumn,bm}

\begin{document}
\title{Fluid breakup during simultaneous two-phase flow \\through a three-dimensional porous medium}

    \author{Sujit S. Datta}
    \affiliation{Department of Physics, Harvard University, Cambridge MA 02138, USA}               \affiliation{Current address: Department of Chemical Engineering, California Institute of Technology, Pasadena, CA 91125}           

    \author{Jean-Baptiste Dupin}
            \affiliation{Department of Physics, Harvard University, Cambridge MA 02138, USA}
            \affiliation{ESPCI ParisTech, Paris, France}

    \author{David A. Weitz}
    \affiliation{Department of Physics, Harvard University, Cambridge MA 02138, USA}
\email{weitz@seas.harvard.edu}

\date{\today}

\begin{abstract}   
We use confocal microscopy to directly visualize the simultaneous flow of both a wetting and a non-wetting fluid through a model three-dimensional (3D) porous medium. We find that, for small flow rates, both fluids flow through unchanging, distinct, connected 3D pathways; in stark contrast, at sufficiently large flow rates, the non-wetting fluid is broken up into discrete ganglia.  By performing experiments over a range of flow rates, using fluids of different viscosities, and with porous media having different geometries, we show that this transition can be characterized by a state diagram that depends on the capillary numbers of both fluids, suggesting that it is controlled by the competition between the viscous forces exerted on the flowing oil and the capillary forces at the pore scale. Our results thus help elucidate the diverse range of behaviors that arise in two-phase flow through a 3D porous medium.

\end{abstract}
\maketitle

\section{Introduction}
Many important technological situations entail the flow of two immiscible fluids through a porous medium; examples include oil recovery, ground water remediation, geological CO$_{2}$ storage, and the operation of trickle bed reactors \cite{bear}. The flow of a {\it single} fluid is typically modeled using Darcy's law, $\Delta p=\mu (Q/A)L/k$; this relates the pressure drop $\Delta p$ across a length $L$ of the porous medium to the volumetric fluid flow rate $Q$ via the fluid viscosity $\mu$, the cross-sectional area of the medium $A$, and the absolute permeability $k$, a constant that depends on the geometry of the pore space. When a second immiscible fluid occupies part of the pore space, it reduces the permeability of the medium to the first fluid; in this case, Darcy's law is typically modified by replacing $k$ with $\kappa k$, where $\kappa<1$ is known as the relative permeability of the first fluid. This parameter can depend sensitively on a bewildering array of factors, such as the proportions of the two fluids, their flow rates, and their viscosities. Characterizing this dependence is particularly important in the oil industry, in which laboratory measurements of $\kappa$ are a critical input to large-scale models of flow within a reservoir. 

Such measurements are frequently performed by simultaneously flowing both fluids through the medium at independently-controlled flow rates, and measuring, for each fluid, the difference in the pressure between two points separated along the length of the medium \cite{honarpour}. Due to the interfacial tension between the two fluids, they are thought to flow through distinct, unchanging, bicontinuous pathways \cite{richards,leverett}. The measured pressure differences are thus interpreted as the pressures dropped in the different fluids, and the relative permeabilities are then calculated using Darcy's law. However, how exactly the two fluids are configured, and how this in turn influences their permeabilities, is poorly understood; indeed, even the validity of the assumption that the fluids flow through connected pathways remains intensely debated \cite{richards,leverett,payatakes,maloy,rice,chatenever}. Experiments on a two-dimensional (2D) porous medium challenge this notion: in some cases, the non-wetting fluid instead breaks up into discrete ganglia, often as small as one pore in size, which are then advected through the medium by the flowing wetting fluid \cite{payatakes,maloy,rice,chatenever}. However, the pore space of a 2D medium is considerably less connected than that of a three-dimensional (3D) medium \cite{sahimi}; this enhances the propensity of the non-wetting fluid to break up in the 2D case \cite{esther}, and thus, the applicability of such experiments to 3D porous media is unclear. Unfortunately, probing the flow in 3D, at pore-scale resolution, is enormously challenging, due to the opacity of the medium; this is exacerbated when multiple fluids are used. Consequently, despite its significant practical importance, an understanding of whether -- and if so, under what conditions -- fluid breakup occurs during simultaneous two-phase flow through a 3D porous medium remains elusive. 

Here, we report the results of confocal microscopy experiments directly visualizing the simultaneous flow of both a wetting and a non-wetting fluid through a 3D porous medium, at pore-scale resolution. For small flow rates, both fluids flow through unchanging, distinct, bicontinuous 3D pathways. At higher flow rates, however, the non-wetting fluid continually breaks up into discrete ganglia; these are then advected through the medium. We propose that the non-wetting fluid breaks up when the sum of the viscous forces exerted by the wetting and the non-wetting fluids exceed the capillary forces at the pore scale. We test this hypothesis by exploring a range of flow rates, using fluids of different viscosities, and porous media having different geometries. Consistent with this picture, we find that the transition between fully connected and broken up flow can be summarized by a state diagram that depends on the capillary numbers of both the wetting and the non-wetting fluids, $Ca_{w}$ and $Ca_{nw}$, respectively. Our results thus help elucidate the diverse range of behaviors that arise in two-phase flow through a 3D porous medium.

 \begin{figure}
\begin{center}
\includegraphics[width=7.5in]{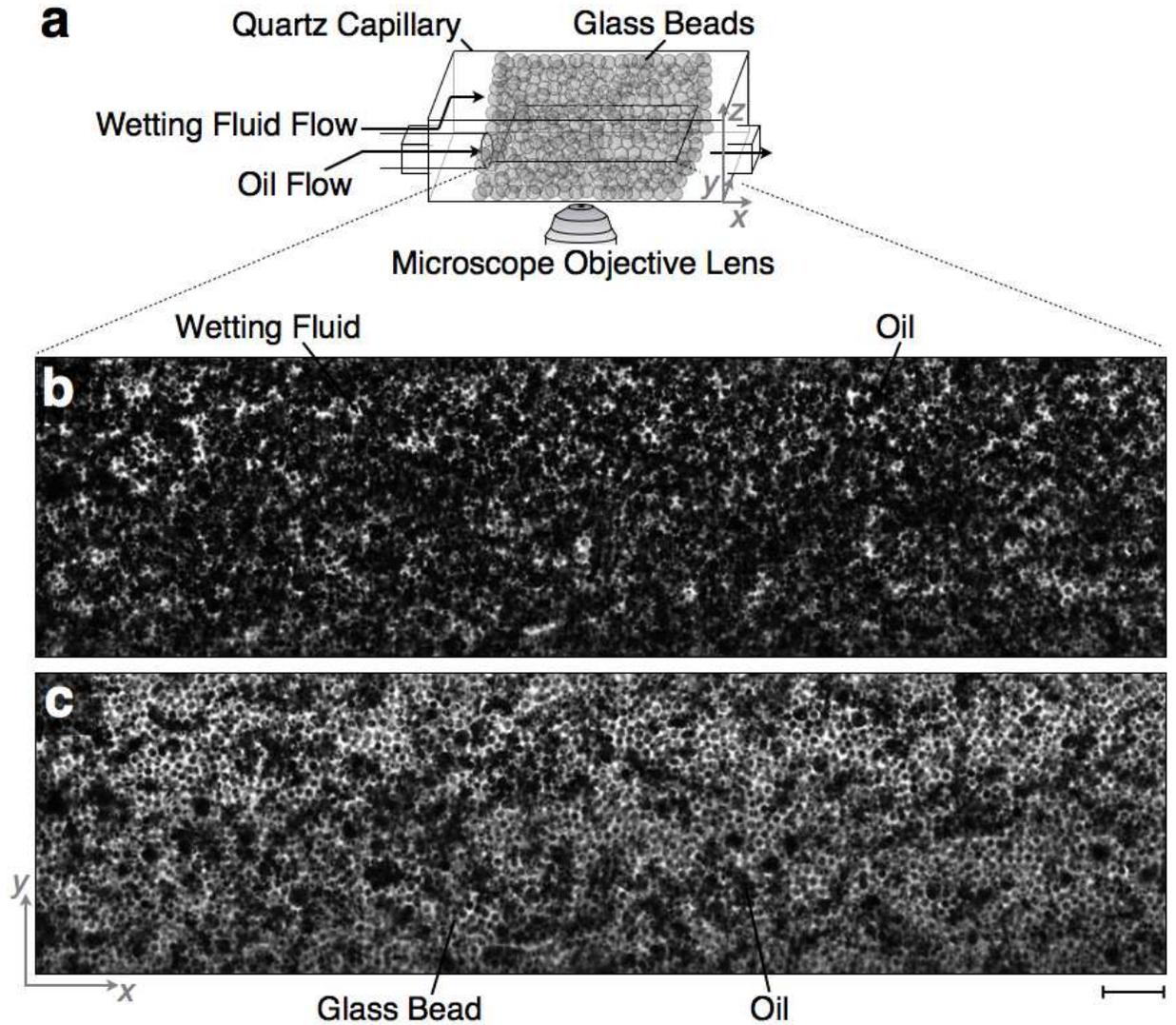}
\caption{(a) Schematic showing porous medium and imaging geometry. The oil flows at a rate $Q_{nw}$ through a circular tube inserted coaxially in a larger square quartz capillary, whereas the wetting fluid flows at a rate $Q_{w}$ via two additional opposing circular tubes (not shown for clarity) positioned at the inlet of the square capillary, through the interstitial space between the oil circular tube and the square capillary. A portion of a 2D confocal section within the medium is shown for (b) $Q_{w}=0.2$ $\mbox{mL~h}^{-1}$ and $Q_{nw}=5.0$ $\mbox{mL~h}^{-1}$, and (c) $Q_{w}=50.0$ $\mbox{mL~h}^{-1}$ and $Q_{nw}=5.0$ $\mbox{mL~h}^{-1}$. Black circles show beads making up the medium, while additional black regions show a section through the flowing oil. Scale bar is 500 $\mu$m long. The oil flows through an unchanging, tortuous, connected 3D pathway in both cases; an example of a 3D reconstruction is shown in Figure 1 of the supplementary materials \cite{supp}. The imposed flow direction is from left to right.} 
\end{center}
\end{figure}

\section{Experimental methodology}
We prepare a rigid 3D porous medium by lightly sintering a dense, disordered packing of hydrophilic glass beads \cite{stokes,plona}, with radii $a=36\pm2~\mu$m, in a thin-walled square quartz capillary of cross-sectional area $A=9$ mm$^{2}$; the typical radius of a pore is thus $a_{p}\approx6~\mu$m \cite{raoush,thompson}. The packing has length $L\approx2$ cm and porosity $\phi\approx0.41$, as measured using confocal microscopy. Scattering of light from the surfaces of the beads typically precludes direct observation of the flow within the medium. We overcome this limitation by formulating fluids whose refractive indices are carefully matched with that of the glass beads \cite{amber,pof,schlumb,prerna}; the wetting fluid is a mixture of 91.4 wt\% dimethyl sulfoxide and 8.6 wt\% water, dyed with fluorescein, while the non-wetting oil is a mixture of aromatic and aliphatic hydrocarbons. The viscosities of the wetting fluid and the oil are $\mu_{w}=2.7~\mbox{mPa~s}$ and $\mu_{nw}=16.8~\mbox{mPa~s}$, respectively. The interfacial tension between the two fluids is $\gamma\approx13~\mbox{mN~m}^{-1}$, as measured using a du No\"{u}y ring. 

To prepare the porous medium, we first evacuate it under vacuum and saturate it with CO$_{2}$ gas, which is soluble in the wetting fluid, thereby minimizing the formation of trapped bubbles. We then saturate the pore space with the dyed wetting fluid; this protocol enables us to visualize the pore structure within the 3D medium using a confocal microscope, as schematized in Figure 1(a). We do this by acquiring optical slices at a fixed $z$ position several bead diameters deep within the medium; each slice is 11 $\mu$m thick along the $z$ axis and spans a lateral area of 912 $\mu$m $\times$ 912 $\mu$m in the $xy$ plane. We repeatedly acquire slices at multiple locations in the $xy$ plane spanning the entire width and length of the medium; by combining these, we obtain a map of the pore structure over the entire extent of the porous medium, and use this map for all following experiments. We identify the glass beads by their contrast with the dyed wetting fluid.

\section{Results and discussion}

To investigate the multiphase flow, we simultaneously flow both the wetting fluid and the non-wetting oil at independently-controlled volumetric flow rates, $Q_{w}=0.2$ $\mbox{mL~h}^{-1}$ and $Q_{nw}=5.0$ $\mbox{mL~h}^{-1}$, respectively. We do this by flowing the oil through a circular tube inserted coaxially well within the square capillary containing the porous medium, and flowing the wetting fluid through the interstitial space between the circular and the square capillaries, via two additional opposing circular tubes positioned at the inlet of the square capillary; this geometry is schematized in Figure 1 of the supplementary materials \cite{supp}. Because the oil is undyed, we identify it by its additional contrast with the dyed wetting fluid in the imaged pore space. The oil initially flows into the medium through a series of abrupt bursts into the pores, remaining connected as it flows; this behavior indicates that a threshold pressure must build up in the oil, to overcome the capillary pressure $\sim\gamma/a_{p}$, before it can invade a pore \cite{amber,pof}. Because the packing of the beads is disordered, the path taken by the oil is tortuous; once it traverses the entire medium, the oil continues to flow through this unchanging, connected 3D pathway, as exemplified in Figure 1(b) and Movie 1 \cite{supp}. We verify that the oil flows through this pathway by monitoring the fluid ejection at the outlet of the porous medium.

 \begin{figure}
\begin{center}
\includegraphics[width=6.4in]{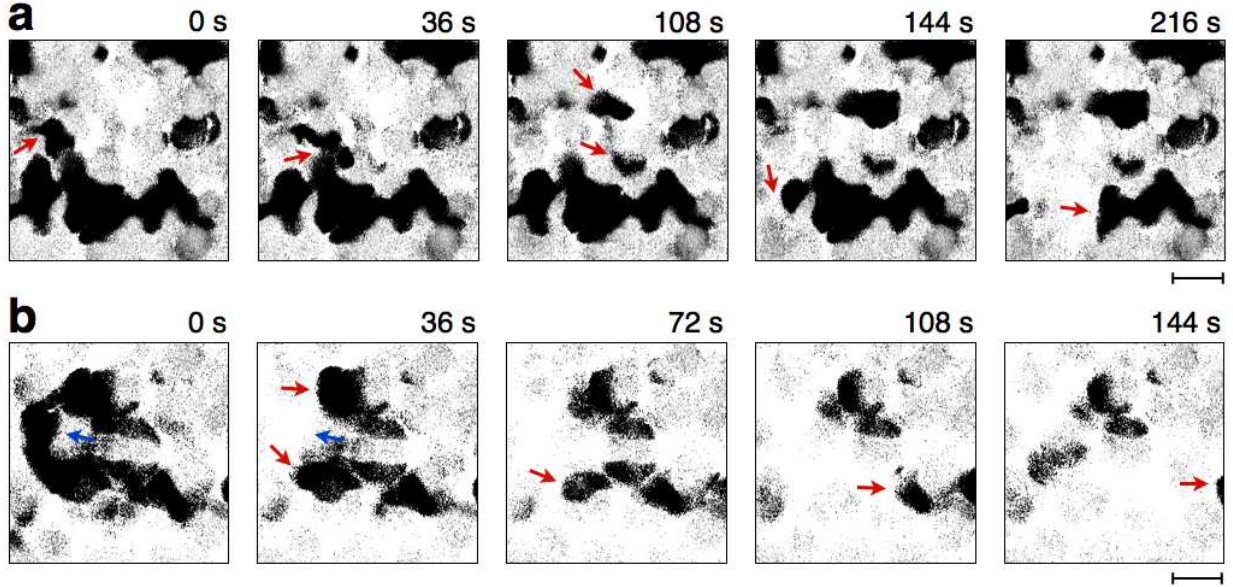}
\caption{Oil breaks up for sufficiently large flow rates. Images show multiple frames, taken at different times, of a single optical slice, with the beads and the pore space subtracted; thus, the dark regions in each frame only show the oil as it flows. The imposed flow direction is from left to right. (a) At $Q_{w}=125.0$ $\mbox{mL~h}^{-1}$ and $Q_{nw}=5.0$ $\mbox{mL~h}^{-1}$, the wetting fluid breaks the oil up into smaller, discrete ganglia. (b) At $Q_{w}=50.0$ $\mbox{mL~h}^{-1}$ and $Q_{nw}=25.0$ $\mbox{mL~h}^{-1}$, the oil breaks up as it flows around the beads forming the porous medium. Labels show time elapsed after first frame. Scale bars are 50 $\mu$m long. Movies are provided in the supplementary materials \cite{supp}. Ganglia are indicated by red right-pointing arrows, while the position of a reference bead is indicated by the blue left-pointing arrow.} 
\end{center}
\end{figure}

To further explore the two-phase flow, we progressively increase the wetting fluid flow rate. For $Q_{w}<125.0$ $\mbox{mL~h}^{-1}$, the oil slightly reconfigures its flow path each time $Q_{w}$ is changed [Figure 1 (b-c)]; it then continues to flow through this new connected 3D pathway. We observe dramatically different behavior at even larger flow rates: instead of simply flowing through a connected 3D pathway, the oil continually breaks up into discrete ganglia, several pores in size, indicated by the red right-pointing arrows in Figure 2(a) and Movies 2-4 \cite{supp}. The ganglia [shown by the red right-pointing arrows] are then advected through the pore space by the flowing wetting fluid. This is a dynamic process \cite{payatakes,maloy}: the oil ganglia continually break up and coalesce, temporarily becoming immobilized at some pore entrances and eventually being pushed through others, ultimately becoming mobilized from the medium. These observations contradict the idea that the oil flows through an unchanging, connected pathway \cite{richards,leverett,honarpour}. 

To investigate the influence of increasing the oil flow rate, we progressively increase $Q_{nw}$, starting at $0.2$ $\mbox{mL~h}^{-1}$, fixing the wetting fluid flow rate at $Q_{w}=25.0$ $\mbox{mL~h}^{-1}$. The oil again flows through an unchanging, connected 3D pathway, similar to those shown in Figure 1(b-c). Similar to the increasing $Q_{w}$ case, the oil slightly reconfigures its flow path each time $Q_{nw}$ is changed; it then continues to flow through this new connected 3D pathway. Intriguingly, at $Q_{nw}=5.0$ $\mbox{mL~h}^{-1}$, sections of this connected 3D pathway intermittently break up into discrete ganglia, several pores in size -- the observed break up process occurs irregularly, punctuated by intervals of continuous flow spanning several seconds to several minutes. At even higher flow rates, we again observe that the oil continually breaks up, without interruption, into discrete ganglia, indicated by the red right-pointing arrows in Figure 2(b) and Movie 5 \cite{supp} -- the blue left-pointing arrow in the Figure indicates the position of a reference bead. Similar to the breakup observed in the case of increasing $Q_{w}$, these ganglia are advected through the pore space by the flowing wetting fluid, and are ultimately mobilized from the medium. Our experiments thus demonstrate that, for sufficiently large fluid flow rates, the two fluids do {\it not} flow through distinct, unchanging, connected 3D pathways, as is typically assumed \cite{richards,leverett,honarpour}; instead, we observe a transition to a state in which the oil is continuously broken up into discrete ganglia.

The first class of transition from fully connected to broken up flow occurs for increasing $Q_{w}$. We observe that, as the oil flows into the porous medium, it is quickly broken up into large ganglia, many pores in size; a representative example is shown in the first frame of Figure 2(a). The wetting fluid pushes part of this ganglion forward [arrows, first two frames of Figure 2(a)], ultimately breaking it off. This process forms additional, smaller ganglia [arrows, third frame of Figure 2(a)]. The ganglia are eventually advected through the pore space by the flowing wetting fluid [arrows, last two frames of Figure 2(a)]. This behavior is reminiscent of drop formation in a microfluidic channel \cite{tjunction1,tjunction2,andy1,crocker}, in which the viscous drag exerted by a flowing wetting fluid tears discrete drops off a coflowing oil stream. Moreover, similar to the case of drop formation, we observe that the size of the oil ganglia formed decreases with increasing $Q_{w}$ [Movies 2-3] \cite{supp}. We thus hypothesize that a similar mechanism drives this transition. We estimate the pore-scale viscous drag force as $\sim\mu_{w}q_{w}a_{p}$, where $q_{w}\equiv Q_{w}/\phi A$ is the average interstitial velocity in the case of single-phase wetting fluid flow. The oil breakup is resisted by the capillary force $\sim\gamma a_{p}$; balancing these two forces yields a criterion for breakup, $Ca_{w}\equiv\mu_{w}q_{w}/\gamma\geq Ca^{*}_{w}$, where $Ca^{*}_{w}$ is a threshold capillary number. 
 
The second class of transition from fully connected to broken up flow occurs for increasing $Q_{nw}$. This behavior is counterintuitive: increasing the flow rate of a non-wetting fluid in a coflowing stream typically {\it impedes} its breakup into drops \cite{andy1,andy2,crocker}. A clue to the physical origin of this transition comes from close inspection of the oil breakup process; a representative example is shown in Figure 2(b) and Movie 5 \cite{supp}.  As the oil flows, it collides with a bead making up the porous medium, indicated by the blue left-pointing arrow in the first two frames of Figure 2(b). It then flows around the entire bead, breaking up into two smaller ganglia in the process [leftmost red right-pointing arrows, second frame of Figure 2(b)]. These smaller ganglia continue to be advected through the pore space by the flowing wetting fluid [arrows, last three frames of Figure 2(b)]. This behavior is reminiscent of the breakup of a non-wetting fluid as it flows past an obstacle in a microfluidic channel \cite{suzie1,suzie2}; in such a situation, for small fluid flow rates, the capillary force causes the non-wetting fluid to squeeze into one of the gaps around the obstacle, moving past it without breaking. By contrast, at sufficiently large flow rates, the viscous force driving the non-wetting fluid pushes it around the entire obstacle, which breaks the fluid up in the process. We hypothesize that a similar mechanism drives this connected-to-broken up transition. We estimate the pore-scale viscous force driving the oil flow as $\sim\mu_{nw}q_{nw}a_{p}$, where $q_{nw}\equiv Q_{nw}/\phi A$ is the average interstitial velocity in the case of single-phase oil fluid flow. The pore-scale capillary force is again given by $\sim\gamma a_{p}$; balancing these two forces yields another criterion for breakup, $Ca_{nw}\equiv\mu_{nw}q_{nw}/\gamma\geq Ca^{*}_{nw}$, where $Ca^{*}_{nw}$ is a threshold capillary number. 
 \begin{figure}
\begin{center}
\includegraphics[width=6.9in]{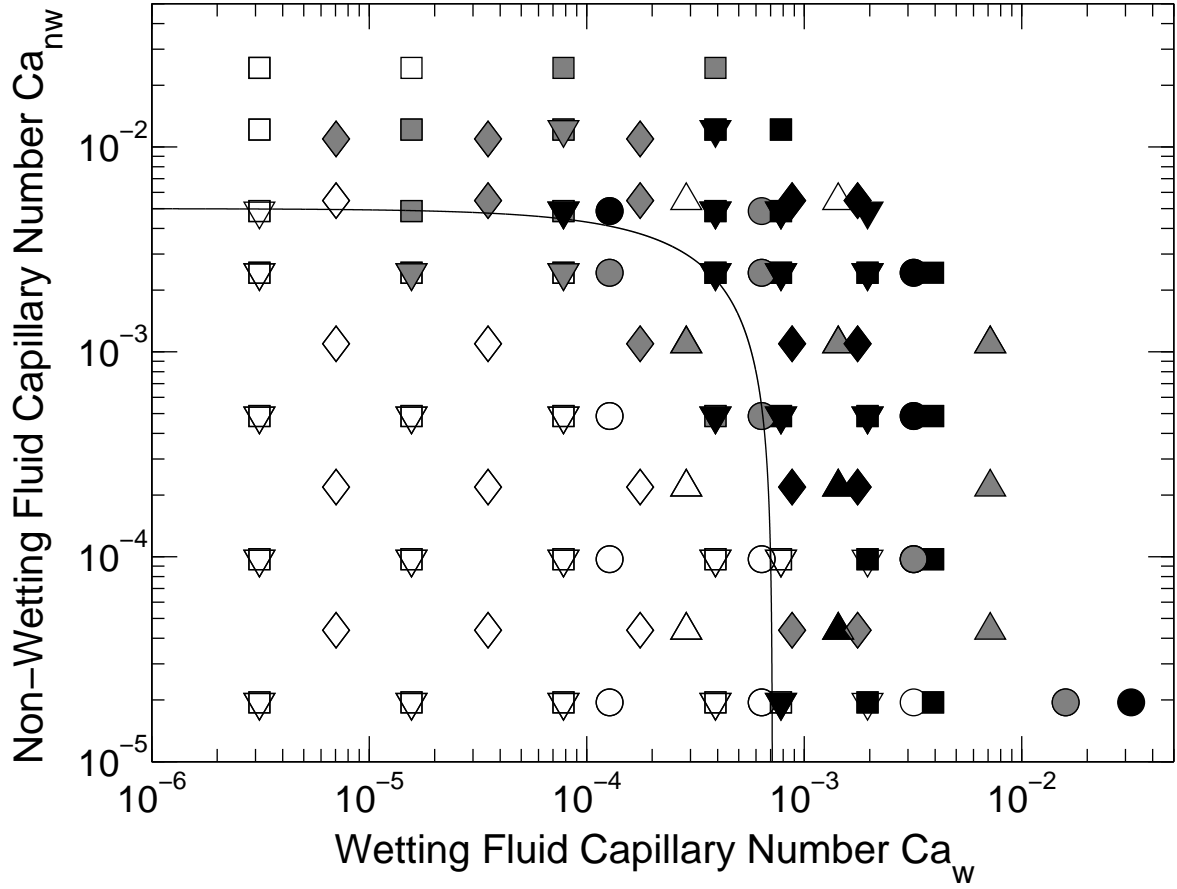}
\caption{State diagram of the transition from fully connected to broken up flow for simultaneous two-phase flow through a 3D porous medium as a function of $Ca_{w}$ and $Ca_{nw}$. Open symbols represent flow of the non-wetting fluid through a connected 3D pathway, grey symbols represent intermittent breakup of this pathway, and black symbols represent continual breakup of the oil into discrete ganglia, which are advected through the pore space. Each symbol shape represents a different viscosity, bead size, or porous medium cross-sectional area. Circles: $\mu_{w}=110.0~\mbox{mPa~s}$, $a=36~\mu$m, $A=9~\mbox{mm}^{2}$. Squares: $\mu_{w}=2.7~\mbox{mPa~s}$, $a=36~\mu$m, $A=9~\mbox{mm}^{2}$. Upward triangles: $\mu_{w}=110.0~\mbox{mPa~s}$, $a=36~\mu$m, $A=4~\mbox{mm}^{2}$. Diamonds: $\mu_{w}=2.7~\mbox{mPa~s}$, $a=36~\mu$m, $A=4~\mbox{mm}^{2}$. Downward triangles: $\mu_{w}=2.7~\mbox{mPa~s}$, $a=60~\mu$m, $A=9~\mbox{mm}^{2}$. } 
\end{center}
\end{figure}

Within the picture presented here, the breakup of the oil occurs for sufficiently large values of $Ca_{w}$ and $Ca_{nw}$. To rigorously test this hypothesis, we perform similar measurements over a broad range of flow rates ranging from 0.2 to 250.0 $\mbox{mL~h}^{-1}$, using different fluids characterized by viscosities ranging from 2 to 110~\mbox{mPa~s}, on different 3D porous media characterized by cross-sectional areas ranging from 4 to 9~\mbox{mm}$^{2}$ and bead radii ranging from 36 to 60$~\mu$m; we thus vary the capillary numbers over the range $Ca_{w}\sim10^{-6}$ to $10^{-1}$ and $Ca_{nw}\sim10^{-5}$ to $10^{-1}$. We summarize our observations using the state diagram shown in Figure 3. In all cases, we observe that the oil flows through an unchanging, connected 3D pathway for small values of $Ca_{w}$ and $Ca_{nw}$, shown by the unfilled symbols in Figure 3; however, as either $Ca_{w}$ or $Ca_{nw}$ increase, this pathway begins to intermittently break up, shown by the grey symbols in Figure 3. At the largest values of $Ca_{w}$ or $Ca_{nw}$, the oil continually breaks up into discrete ganglia, as shown by the black symbols in Figure 3. All the data appear to collapse, supporting the validity of our picture \cite{history}. 

The first class of connected-to-broken up transition occurs when $Ca_{w}$ exceeds a threshold value, $Ca_{w}^{*}$; we use our measurements to estimate this value, $Ca_{w}^{*}=7\times10^{-4}$, shown on the abscissa in Figure 3. Interestingly, this value is smaller than the threshold for drop formation in a microfluidic junction, $Ca_{w}^{*}\sim10^{-2}$ \cite{tjunction1,tjunction2}. We speculate that this difference reflects the disordered packing of the beads. Indeed, direct visualization using confocal microscopy shows that, even in the case of single-phase flow, the velocity of the flowing wetting fluid can be over an order of magnitude larger than $q_{w}$, due to the tortuosity of the pore space \cite{fluctuations}. Moreover, in our experiments, the co-flowing oil phase occupies a large portion of the pore space; this increases the average interestitial velocity, and can impart even more variability to the wetting fluid flow. We thus expect that the pore-scale viscous drag forces can be significantly larger than the value we use in our simple estimate, $\mu_{w}q_{w}a$; this leads to a corresponding reduction in $Ca_{w}^{*}$. Adapting our imaging approach to directly measure the local flow velocities will thus be an important extension of the work described here. The second class of connected-to-broken up transition occurs when $Ca_{nw}$ exceeds a threshold value, $Ca_{nw}^{*}$; we use our measurements to estimate this value, $Ca_{nw}^{*}=5\times10^{-3}$, shown on the ordinate in Figure 3. Interestingly, this value is comparable to the threshold for the breakup of a non-wetting fluid as it flows around an obstacle in a microfluidic channel \cite{suzie1,suzie2,viscnote}. 

We note that the observed transition between fully connected and broken up flow may depend on a number of different physical forces, including the fluid inertia and gravity -- for simplicity, we only consider viscous and capillary forces here. Motivated by the collapse of the data in Figure 3, we suggest that the connected-to-broken up transition occurs when the sum of the viscous forces exerted on the flowing oil exceeds capillary forces at the pore scale; this can be summarized as $Ca_{w}/Ca_{w}^{*}+Ca_{nw}/Ca_{nw}^{*}\sim1$. We find good agreement between this relation and our experimental measurements, as shown by the curve in Figure 3. This provides further support for the validity of the picture presented here.

\section{Conclusions}

Our experiments reveal that, when both a wetting and a non-wetting fluid are forced to flow through a porous medium simultaneously, the non-wetting fluid may not flow through an unchanging, connected pathway, as is typically assumed. Instead, it can continually break up into discrete ganglia, which are then advected through the pore space by the wetting fluid. This may complicate the measurement of a pressure drop in the non-wetting fluid, and consequently, the use of Darcy's law to calculate the relative permeability. Indeed, we find that the fluid breakup can occur at capillary numbers comparable to those used in some permeability measurements \cite{richards,leverett,blunt,fulcher}.

\acknowledgements 
It is a pleasure to acknowledge E. Amstad, M. P. Brenner, T. S. Ramakrishnan, and H. A. Stone for stimulating discussions, H. Chiang for experimental assistance, and the anonymous referees for invaluable feedback. This work was supported by the NSF (DMR-1006546), the Harvard MRSEC (DMR-0820484), and the Advanced Energy Consortium (http://www.beg.utexas.edu/aec/), whose member companies include BP America Inc., BG Group, Petrobras, Schlumberger, Shell, and Total. SSD acknowledges funding from ConocoPhillips.


\begin{thebibliography}{55}
\bibitem[{history()}]{history}
\bibinfo{note}{Intriguingly, we do not observe any strong dependence on the flow history, consistent with other work \cite{maloy2,mri}.}
\bibitem[{viscnote()}]{viscnote}
\bibinfo{note}{Surprisingly, unlike the case of breakup around an obstacle in a microfluidic channel, we do not observe any strong dependence of $Ca_{nw}^{*}$ on the viscosity ratio of the fluids used.}
\end{thebibliography}
\end{document}